\documentclass[aip,jcp,amsmath,amssymb,floatfix,reprint,twocolumn]{revtex4-2}
\usepackage{color,soul}
\usepackage{graphicx}
\usepackage{amsmath, amsthm, amssymb,mathtools}
\usepackage{amsthm}
\usepackage{multirow}
\usepackage[normalem]{ulem}
\usepackage{soul}
\usepackage{hyperref}

\usepackage{bm}
\usepackage{bigdelim}
\usepackage{array}
\usepackage{float}
\usepackage{subcaption}
\usepackage[table]{xcolor}
\arrayrulecolor{gray}
\newcommand{\mycomment}[1]{}

\newcommand{\be}{\begin{equation}}
\newcommand{\ee}{\end{equation}}
\newcommand{\bea}{\begin{eqnarray}}
\newcommand{\eea}{\end{eqnarray}}
\makeatletter

\newcommand{\Rmnum}[1]{\expandafter\@slowromancap\romannumeral #1@}
\makeatother\usepackage{array, makecell}

\DeclareMathOperator{\arcsinh}{arcsinh}

\begin{document}

\title{Emerging universality classes in thermally-assisted activation of interacting diffusive systems: A perturbative hydrodynamic approach}

\author{Vishwajeet Kumar}
\affiliation{The Institute of Mathematical Sciences, C.I.T. Campus, Taramani, Chennai 600113, India}
\affiliation{Homi Bhabha National Institute, Training School Complex, Anushakti Nagar, Mumbai 400094, India}
\author{Arnab Pal} 
\email{arnabpal@imsc.res.in}
\affiliation{The Institute of Mathematical Sciences, C.I.T. Campus, Taramani, Chennai 600113, India}
\affiliation{Homi Bhabha National Institute, Training School Complex, Anushakti Nagar, Mumbai 400094, India}
\author{Ohad Shpielberg}
\email{ohads@sci.haifa.ac.il}
\affiliation{Department of Mathematics and Physics, University of Haifa at Oranim, Kiryat Tivon 3600600, Israel}
\affiliation{Haifa Research Center for Theoretical Physics and Astrophysics, University of Haifa, Abba Khoushy Ave 199, Haifa 3498838, Israel}


\begin{abstract}
Thermal activation of a particle from a deep potential trap follows the Arrhenius law. Recently, this result was generalized for interacting diffusive particles in the trap, revealing two universality classes -- the Arrhenius class and the excluded volume class. The result was demonstrated with the aid of numerical analysis. Here, we present a perturbative hydrodynamic approach to analytically validate the existence and range of validity for the two universality classes.   
\end{abstract}

\maketitle

\noindent



\section{Introduction}
\label{intro}

%

Thermally-assisted activation processes represent a pervasive phenomenon with wide-ranging applications across various scientific and technological domains. At the core of understanding these processes lies Kramers theory, which provides a comprehensive and fundamental theoretical framework. This theory elucidates the underlying principles of thermally-assisted activation, e.g. in chemical reaction rates, laser pumping and magnetic resonance, and famously for description of folding and unfolding of proteins and nucleic acids \cite{lapolla2020single,hanggi1990reaction,pollak2005reaction}.

When the thermal energy $D_0= k_B T$ is weak compared with the  barrier crossing energy $\Delta U$, the Kramers theory leads to the Arrhenius law (AL) -- a seminal result of Physical Chemistry. See \cite{hanggi1990reaction} for a historical review and \cite{mel1986theory} for exact results. The AL states that the escape rate to overcome the barrier is given by 
\begin{align}
\Phi = \tau^{-1}_0 \rm{e}^{-\Delta U / D_0}.
\label{AL-main}
\end{align}
 Here, the pre-exponential $\tau_0 $ is a microscopic time scale, which is polynomial in $\Delta U$. The salient aspect of the AL is the exponential part; It is independent of the details of the barrier. Therefore, the AL is particularly appealing as the  barrier energy can be inferred by measuring the escape times at different temperatures. 

Due to the success of the AL and even after a century of scientific study, current research continues to evolve our understanding of thermally-assisted barrier crossing, extending beyond the Kramers theory.
For example, the generality and universality of the AL has been put to the test in systems with long range interaction  \cite{mukamel2005breaking,saadat2023lifetime}, multiple meta-stable states  \cite{thorneywork2020direct,chupeau2020optimizing}, colored noise \cite{del2023escape,dubkov2023enhancement}, and recently also tested in active systems \cite{woillez2019activated,woillez2020nonlocal}. Surprisingly enough, the problem of AL for interacting systems received little attention over the years \cite{langer1969statistical,hennig2015cooperative,sebastian2000kramers}. In what follows, we postulate a setup, where \eqref{AL-main} cannot hold.

Consider a system of finite size colloids, restricted to move in a $1D$ tube, and trapped by an external potential. If the density of colloids in the trap is finite, they cannot all occupy the bottom of the potential as some of the colloids are pushed away from the bottom. Hence, an effective activation energy, smaller than $\Delta U$, may be considered. Naively, this effective activation energy can depend on the inter-colloid interaction and the interaction of the colloid with the medium. Clearly, it also depends on the density of colloids in the trap, ultimately  suggesting a violation of the AL \eqref{AL-main}. 
This simple setup rightly shows the key role of interactions in the escape problem of AL and why a new formulation is very much needed to generalize the AL for interacting systems.



In a recent publication \cite{kumar2023arrhenius}, we have demonstrated that there exists a generalized AL for a large class of diffusive interacting systems within the framework of the macroscopic fluctuation theory (MFT) and with the aid of extensive numerics. Following \cite{kumar2023arrhenius}, we derived 
\begin{equation}
    \Phi \asymp \rm{e}^{-\Delta U g(\overline{\rho}_0 )  / D_0  } ,
    \label{gen-AL}
\end{equation}
which is the generalized AL. Here, $\overline{\rho}_0$ is the mean density of the particles in the trap and $\Delta U g$ is an effective activation energy. Note that $g$ depends both on the 
density $\overline{\rho}_0$ and on the inter-particle interactions. As will be shown later, $ g = 1-  U_{top}/\Delta U $ where $ U_{top}$ is the highest energy a particle can attain in the minimum energy configuration. 

The central goal of this work is to provide an analytical derivation for the generalized AL (\ref{gen-AL}) using a perturbative hydrodynamic approach. In particular, we present a generic scheme to derive the generalized AL using perturbation theory and then apply the same to several lattice gas models of interest. 
A complementary goal of this endeavor, essential for accomplishing our primary goal, is to present a non-expert and practical guide for applying the MFT to analyze escape rates in the presence of a potential trap (whether $\Delta U \gg D_0$ or not). 

The salient features of this work can be summarized in the following. First, the exponential form of the AL is still preserved in \eqref{gen-AL}. 
Second, we identify two universality classes emanating from the generalized AL. For interacting particles with excluded volume, we find $g<1$ which means that the escape rate is enhanced due to the interaction among particles. On the other hand, for interacting particles without excluded volume, we find $g=1$ which essentially means that the canonical AL is preserved. Importantly, universality here means that $g$ is independent of the dynamics of the interacting particles. 
In this work, along with the derivation of the generalized AL, we also unravel this universality and discuss within a class of stochastic interacting systems.

\begin{figure*}
  \centering
  \begin{subfigure}[b]{0.4\linewidth}
    \centering
    \includegraphics[width=\linewidth]{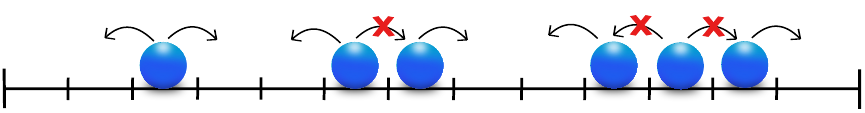}
    \caption{The SEP}
    \label{subfig:a}
  \end{subfigure}
  \hspace{1cm} 
  \begin{subfigure}[b]{0.4\linewidth}
    \centering
    \includegraphics[width=\linewidth]{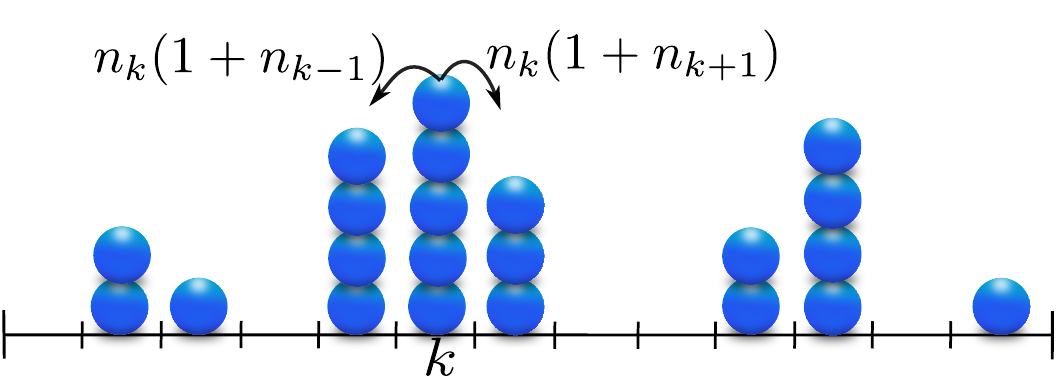}
    \caption{The SIP}
    \label{subfig:b}
  \end{subfigure}

  \vspace{1cm} 

  \begin{subfigure}[b]{0.4\linewidth}
    \centering
    \includegraphics[width=\linewidth]{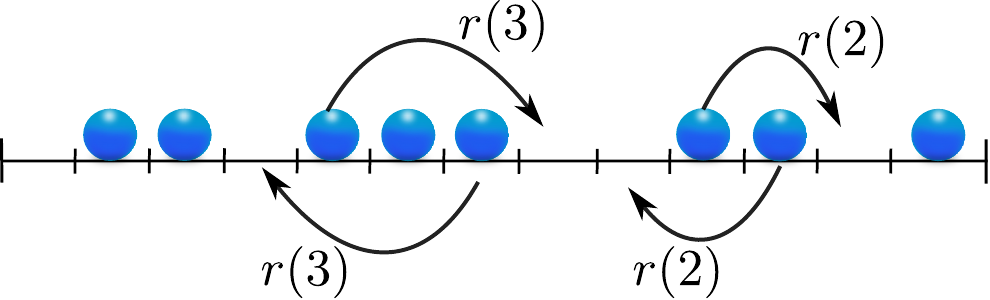}
    \caption{The strong particles model}
    \label{subfig:c}
  \end{subfigure}
  \hspace{1cm} 
  \begin{subfigure}[b]{0.4\linewidth}
    \centering
    \includegraphics[width=\linewidth]{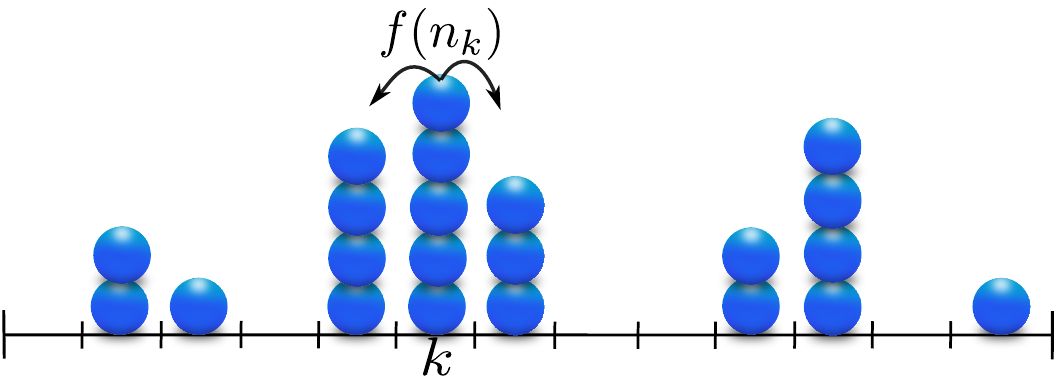}
    \caption{The ZRP}
    \label{subfig:d}
  \end{subfigure}
  \caption{The microscopic dynamics of (a) the SEP (b) the SIP (c) the strong particles model (d) the ZRP. The dynamics of the processes is given with vanishing potential for simplicity of presentation.  
  }
  \label{fig:main}
\end{figure*}

The outline for the rest of the paper is as follows. 
In Sec.~\ref{sec:Kramers Setup}, we recast escape problem in the Kramers setup as the ground state energy of an imaginary-time Schr\"{o}dinger equation. Later on, using the MFT, we will generalize the resulting 
eigenvalue problem to the case of interacting systems. In particular, we will identify that in the non-interacting case, we recover the results of Sec.~\ref{sec:Kramers Setup}. In Sec.~\ref{sec:MFT} the MFT is introduced as a nonequilibrium theory of diffusive systems. Sec.~\ref{sec:Survival} formulates the escape problem in diffusive systems within the framework of the MFT. The formalism is benchmarked by showing that for non-interacting particles, we recover the AL of Sec.~\ref{sec:Kramers Setup}. 
In Sec.~\ref{sec: close to equilibrium approach }, we finally turn to prove the generalized AL presented above. The essence of the proof is presented in Sec.~\ref{subsec: proof essence} for arbitrary diffusive dynamics. The rest of the section serves to apply the method to several eminent interacting systems.
Sec.\ref{sec:discussion} serves to summarise the work and discuss future perspectives. 

\section{The eigenvalue method applied to the Kramers problem}
\label{sec:Kramers Setup}
In this Section, we revisit the Kramers setup and recast the problem of finding the escape rate of a single particle to that of finding the ground state energy of an imaginary-time Schr\"{o}dinger equation. This will prove useful later on, in benchmarking the hydrodynamic approach. To that end, let us recall the Kramers setup accounting for the single particle case.

Consider an over-damped particle, coupled to a thermal bath at temperature $T$, and under the influence of an asymmetric double well potential. The particle is initially located in the meta-stable state at the higher potential minimum. We are interested in the mean escape time for the particle to jump over the energy barrier and relax to the lower potential minimum. For weak thermal noise, there is a separation of time scales. The time for the particle to ascend the potential barrier is significantly longer than the time it takes the particle to descend from the barrier peak towards the lower potential minimum. Therefore, the escape time can be approximated to be just the time it takes to ascend the potential barrier, which follows the AL. 

Following the above premise, we consider the potential $U(X)$ for $X\in \left[0,L\right] $. At $X=0$ the potential vanishes (the higher minimum) and we fix $U(0)=0$ and instate a reflective boundary condition. At $X=L$, the peak of the potential barrier, we fix $U(L)=\Delta U$ and instate an absorbing boundary condition.

The standard way to capture the Kramers setup is via the Fokker-Planck equation
\begin{equation}
    \label{eq: FP equation}
    \partial_t P = D_0 \partial_{XX} P + \partial_X (\partial_X U P), 
\end{equation}
where $P(X,t)$ is the probability to observe the particle at position $X$ and time $t$. The Fokker-Planck equation \eqref{eq: FP equation} corresponds to the boundary conditions 
\begin{equation}
    \label{eq: FP BC}
    \partial_X P + P \partial_X U |_{X=0} = P|_{X=1} =0.   
\end{equation}
Notice that due to the absorbing boundary, the probability is not conserved \cite{gardiner1985handbook,risken1996fokker}. Therefore, it is only natural to consider the
survival probability $S(t)$ -- the probability that after time $t$, the particle is still trapped. By definition,
\begin{equation}\label{Survival Probability}
    S(t) = \int^L _0  dX \, P(X,t) . 
\end{equation}
with
$S(t=0)=1$ and $S(t\rightarrow \infty) =0 $. At large times, significantly larger than the diffusive time $L^2/D_0$, the survival probability is expected to follow a large deviation form $S(t)\asymp \rm{e}^{-\Phi t}$, where we recall that $\Phi$ is the escape rate, given by the AL. 


An explicit solution for the survival probability in the presence of an arbitrary potential $U(X)$ does not exist. However, several methods exist to capture the large time behavior. One such way is to convert the Fokker-Planck equation into an imaginary-time Schr\"{o}dinger equation \cite{gardiner1985handbook}. This can be done by considering the transformation $P(X,t) = \rm{e}^{-U(X)/2D_0} \psi(X,t)$ such that 
$\psi(X,t)$ satisfies the Schr\"{o}dinger equation. Namely
\begin{eqnarray}\label{Schrodinger}
    -\partial_t \psi &=& \hat{H}_{FP} \psi ,
    \\ \nonumber 
    \hat{H}_{FP} / D_0   &=& - \partial_{XX}  + \frac{1}{4}\left(\frac{\partial_X U}{D_0} \right)^2 + \frac{\partial_{XX} U}{2D_0} .
\end{eqnarray}
Quantum-mechanically, the presence of reflecting and absorbing boundaries implies that the particle is bounded. This in turn implies that the energy spectrum of the system would be discrete \footnote{We assume here that $U$ is smooth and bounded.}. Hence, we can now write,
\begin{equation}
    \psi(X,t)=\sum_n C_n \phi_n(X)e^{-E_n t},
\end{equation}
where, $E_n$ is the $n$-th eigenvalue of the Hamiltonian $\hat{H}_{FP}$ in Eq.(\ref{Schrodinger}) and $\phi_n$ is the corresponding eigenvector.

The transformation to the quantum problem is useful when considering the large time limit as it is easy to see that $\psi \sim \rm{e}^{- E_0 t }$, where $E_0$ is the ground state energy of $\hat{H}_{FP}$. Translating the same to the survival probability  [Eq.(\ref{Survival Probability})], one should have, in the large time limit, 
\begin{align}
 S(t) \asymp \rm{e}^{-t  E_0},
\end{align}
which allows us to identify $E_0 = \Phi$. Moreover, we notice that initial conditions hardly effect the long time behaviour \footnote{With the obvious exception that the initial condition  $\psi(x,t=0)$ is almost completely orthogonal to the ground state. Or physically, the particle starts close to the absorbing boundary.}. The argument can be extended to \textit{$M$ non-interacting particles} such that
\begin{equation}
\label{eq:survival prob single particle}
    S(t) \asymp \rm{e}^{- M  E_0 t }, 
\end{equation}
from which one can identify that $\Phi =  M E_0$ for $M$ non-interacting particles, where $E_0$ is the ground state energy of the single particle Hamiltonian $\hat{H}_{FP}$.

At this point it is important to note that finding $E_0$ is non-trivial for an arbitrary potential. Inferring the universal nature of $E_0$, i.e. the AL, is better carried out with different strategies, well known in the literature \cite{gardiner1985handbook,risken1996fokker}. We postpone divulging one such method to Sec.~\ref{sec: close to equilibrium approach }.  In what follows, we present a brief primer on the nonequilibrium hydrodynamic theory namely the Macroscopic Fluctuation Theory (Sec.~\ref{sec:MFT}) that will be the key formalism to compute the survival probability for interacting particles (Sec.~\ref{sec:Survival}).
In particular,  we recast $\Phi$ as a minimization problem. For non-interacting particles, we recover the solution as the ground state energy of  $H_{FP}$, benchmarking the MFT formalism.   



\section{Introduction to the Macroscopic fluctuation theory \label{sec:MFT} }

Since the early 2000' the MFT has been rigorously verified as a coarse grained theory for nonequilibrium diffusive systems. Here, we will forgo a rigorous presentation of the MFT in favor of a more focused presentation. The presentation aims to familiarize the reader with the tools needed to obtain the survival probability of interacting diffusive systems subjected to a potential. For extensive reviews on the MFT, see \cite{MFT15,Derrida07}.

\begin{figure}[b]
    \centering
    \includegraphics[scale=0.55]{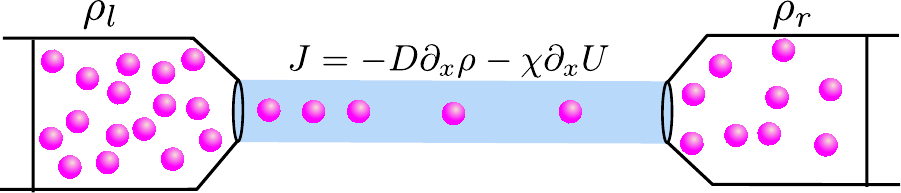}
    \caption{Fick's law for a one dimensional boundary driven system. The boundaries are coupled to particle reservoirs at densities $\rho_l, \rho_r$ correspondingly.  }
    \label{fig:Ficks law}
\end{figure}

For the sake of our presentation, we restrict ourselves to a system of interacting particles in $1D$, with $X\in \left[0,L \right]$. The system is coupled at $X=0,L$ correspondingly to two particle reservoirs with densities $\rho_l , \rho_r$ and both at the fixed temperature $T$. The diffusivity of the system is encapsulated by Fick's law of the current $J = -D \partial_x \rho - \chi \partial_x U$. Here $x= X/L$ is a rescaled length coordinate, $D$ and $\chi$ are the diffusion and mobility that generally depend on the local density. Fick's law implies that the current stems from both a density gradient and from an external forcing due to the potential gradient $\partial_x U$ \footnote{Of course, not all forces are gradient forces. This is of little interest in the present context.  }. See Fig.~\ref{fig:Ficks law} 

To complete the steady state analysis, we assume that interactions conserve the particle number in the bulk. Namely, the dynamics introduces a continuity equation $\partial_t \rho = -\partial_x J $, where particles can flow into/out of the system only at the boundaries.  

Notice that Fick's law, together with the continuity equation imply the diffusion equation. Therefore, we have a deterministic description of the evolution of the system. Yet, it is clear that thermal fluctuations, which are of prime interest in this work, give rise to a stochastic description. We can then consider the fluctuating hydrodynamics 
\begin{eqnarray}
\label{eq:fluc hydro}
    \partial_t \rho &=& - \partial_x j, \\ \nonumber 
    j &=& J(\rho) + \sqrt{2D_0 \chi/L} \, \xi(x,t), 
\end{eqnarray}
where $\xi$ satisfies $\langle \xi(x,t)\rangle =0 $ and $\langle \xi(x,t)\xi(x',t')\rangle = \delta(x,x')\delta(t,t')$. Quite surprisingly, the strength of the noise is related to the mobility, as can be inferred from the Gallavotti-Cohen fluctuation relation \cite{Derrida07}. The fluctuating hydrodynamics in Eq.~\eqref{eq:fluc hydro} provides a full description of the MFT. We notice that the entire dynamics is completely captured by $D$ and $\chi$ alone. This compact description demonstrates the power and simplicity of the MFT. For our purposes, it would be better to recast the fluctuating hydrodynamics using a statistical field theory approach. Employing the Martin-Siggia-Rose formalism \cite{martin1973statistical,kamenev2023field,altland2010condensed}, we find that the probability to observe the fluctuation $\lbrace \rho,j \rbrace $ during the time window $\left[0,t\right]$ is given by the fundamental formula of the MFT 
\begin{eqnarray}
\label{eq:fundamental formula}
    \rm{Prob} \left[ \lbrace \rho,j \rbrace  \right]  & \asymp
 & \rm{e}^{-\frac{L}{4D_0} \int dx dt \, \frac{(j-J(\rho) )^2 }{\chi(\rho)} } , 
\end{eqnarray}
where $\asymp
$ indicates possible logarithmic corrections, and the continuity equation is implicitly assumed. The MFT is a coarse grained theory and hence large systems are assumed. The fundamental formula Eq.~\eqref{eq:fundamental formula} reveals that calculations are dominated by the saddle point due to the large $L$ limit we consider. This fact helped facilitating a large number of results within the MFT framework. From calculations of full counting statistics \cite{bertini2005current,mallick2022exact,bettelheim2022inverse,akkermans2013universal,imparato2009equilibriumlike,hurtado2014thermodynamics}, nonequilibrium correlations \cite{bertini2009towards}, first passage problems \cite{meerson2014survival,agranov2018narrow} and  more \cite{aminov2015fluctuation,krapivsky2014large,grabsch2023joint,shpielberg2019imitating}. At this point we have presented the necessary tools to take head on the problem of finding survival probability of diffusive interacting systems in a potential trap.


\section{The survival probability for interacting particle systems in a trap  }
\label{sec:Survival}

In this section, we finally employ the MFT in order to derive the survival probability of interacting particles in a potential trap. After deriving the formal solution, we focus on the case of non-interacting particles, showing it matches the solution presented in Sec.~\ref{sec:Kramers Setup}.


Let us stress the setup in the language of the MFT. We consider an interacting system, defined by $D,\chi$. The particles are trapped in the potential $U(x)$. The potential global minimum is fixed to zero at $x=0$ and its global maximum lies at  $U(x=1)= \Delta U$. To capture the survival probability of $M$ particles, we impose the restriction  
\begin{equation}
\label{eq:mass in the trap}
\int dx \,  \rho(x,t') = M/L = \overline{\rho}_0    ,
\end{equation}
 for $t'\in \left[0,t\right]$. The  boundary conditions, similarly to Sec.~\ref{sec:Kramers Setup} are reflecting at $x=0$ and absorbing at $x=1$, i.e.  
\begin{equation}
\label{eq:rho BCs}
    J(\rho)|_{x=0} = \rho|_{x=1} =0. 
\end{equation}


The survival probability is the conditional sum over all the probabilities satisfying the boundary conditions \eqref{eq:rho BCs} and the conservation of mass in the trap \eqref{eq:mass in the trap}.  Recall that for the non-interacting case the initial condition is generally immaterial. This is assumed to be valid in the interacting case as well.

The conditional sum is dominated by the saddle point solution, also known as an instanton solution or the optimal fluctuation method. Thus, finding the survival probability in the MFT is recast as the minimization problem 
\begin{equation}
\label{eq:formal minimization}
    S(t) \asymp \rm{e}^{ - L \min_{\rho,j} \int dx dt  \, \frac{(j-J)^2}{4D_0 \chi}  }, 
\end{equation}
subjected to the boundary conditions \eqref{eq:rho BCs}, an initial density profile, the continuity equation, and particle mass $M$ throughout the process \eqref{eq:mass in the trap}.

The saddle point takes us from a hard conditional summation problem, to a simplified minimization problem. Yet, the problem at hand is still formidable. To move forward, it is necessary to simplify the problem further. To that end, let us posit that the solution to the minimization problem leads to a time-independent density profile. To be more precise, there might exist a short time where the initial density profile modifies towards the optimal solution, after which it becomes time-independent. The contribution of this short-time modification to the survival probability at large times is inconsequential and we will treat only the time-independent part of the density profile minimization problem.   

This assumption reduces the complexity of the minimization problem. For a time-independent density in a $1D$ system, the current is constant. The boundary condition at $x=0$ implies that $j$ identically vanishes. Hence, we can write  
\begin{eqnarray}
\label{eq: simplified survival probability}
    S(t)  & \asymp &\rm{e}^{- \frac{L t}{4D_0} \min_{\rho(x)}  \int dx \, \mathcal{L} }, \\ \nonumber 
   \rm{with }\,  \mathcal{L} &=& \frac{J^2 }{\chi } -4\Lambda (\rho - \overline{\rho}_0), 
\end{eqnarray}
and subjected to the boundary conditions \eqref{eq:rho BCs}. Here, we have introduced the Lagrange multiplier $\Lambda $ to account for the mass $M = L \overline{\rho}_0$ in \eqref{eq:mass in the trap}.  

Assuming the density profile is time-independent has been suggested in the context of finding the full counting statistics \cite{bodineau2004current}. This assumption is known as the additivity principle and it proved successful in calculating large deviations within the framework of the MFT and similar theories 
\cite{meerson2016large,hurtado2009test,Shpielberg2016}. Here, we can show that for $M\rightarrow 0 $, the additivity principle is always valid. A breakdown of the additivity principle constitutes a dynamical phase transition. These have been studied in detail \cite{agranov2023tricritical,shpielberg2018universality,shpielberg2017geometrical,shpielberg2017numerical,baek2017dynamical,appert2008universal}. Addressing the possibility of dynamical phase transitions in the context of the AL will be carried out in a subsequent publication \cite{Kumar2024Nonmonotone}.


To find the escape rate from \eqref{eq: simplified survival probability}, one can notice that we have to solve an Euler-Lagrange equation. Before doing so, it is useful to introduce the $F$ transformation, previously used in \cite{agranov2018narrow,meerson2014survival},
\begin{equation}
\label{eq:F transformation}
    F[\rho] = \int^\rho _0  dz \frac{D(z)}{2\sqrt{\chi(z) }}.
\end{equation}
This leads to writing the survival probability as $S(t) \asymp \rm{e}^{-t \Phi}$, where 
\begin{eqnarray}
\label{eq:mini Phi F}
    \Phi &=&  D_0 L  \int dx \,  J^2 _F, \\ \nonumber
    J_F &=&  \partial_x F + \frac{1}{2}\chi^{1/2}(F)\partial_x U / D_0 . 
\end{eqnarray}
Here, $F$ satisfies the Euler-Lagrange equation and the boundary conditions inherited from \eqref{eq:rho BCs}:
\begin{eqnarray}
\label{eq:EL equation F}
    -\partial_{xx}F + \frac{\chi' }{8D^2 _0}(\partial_x U)^2 - \frac{\sqrt{\chi}}{2 D_0}  \partial_{xx}U &=& \Lambda \frac{\sqrt{\chi}}{D}, 
    \\ \nonumber 
    J_F |_{x=0} = F|_{x=1} = 0 .  && 
\end{eqnarray}
We have defined  $\chi' = \partial_F \chi $ in \eqref{eq:EL equation F}. 
Solving the minimization problem \eqref{eq:mini Phi F} together with \eqref{eq:EL equation F} is typically non-trivial analytically. However, numerically it was already addressed using a shooting method \cite{kumar2023arrhenius}. Let us first demonstrate the problem for the case of non-interacting systems. The goal is to show that the MFT solution matches that of Sec.~\ref{sec:Kramers Setup}, thus benchmarking the MFT method.


For non-interacting particles, $D=1,\chi= \rho$. The $F$ transformation \eqref{eq:F transformation} leads to $\rho =F^2$. With that in mind, we find \eqref{eq:EL equation F} becomes 
\begin{equation}
\label{eq:Hfp eigenvalue}
    \hat{H}_{FP} F = \Lambda F, 
\end{equation}
with the boundary conditions \eqref{eq: FP BC}. Namely, $\Lambda$ is the ground state energy of the Fokker-Planck Hamiltonian of Sec.~\ref{sec:Kramers Setup}. Moreover, using the boundary conditions we find from \eqref{eq:mini Phi F}  and \eqref{eq:Hfp eigenvalue}   that $\Phi = M \Lambda $ \cite{kumar2023arrhenius}. To ensure minimization, $\Lambda = E_0$ the ground state energy of the Hamiltonian $\hat{H}_{FP}$. Hence, the MFT result coincides for non-interacting particles with the survival probability calculated in Sec.~\ref{sec:Kramers Setup}, benchmarking the MFT formalism.

Despite not being able to solve the minimization problem analytically for arbitrary potentials, the MFT leads to the AL for non-interacting particles, as expected. Now, for non-interacting systems, the escape rate is hard to obtain, except for a few handpicked potentials. However, we note that at the limit of $\overline{\rho}_0 \rightarrow 0$, interactions becomes negligible, and we recover the non-interacting limit and hence the AL.

In the next section, we tackle the minimization problem with interactions and at finite density $\overline{\rho}_0$. We circumvent attempting an exact solution of the Euler-Lagrange equation  \eqref{eq:EL equation F}, and instead propose a perturbative approach centered around thermal equilibrium.


\section{A perturbation theory close to equilibrium
\label{sec: close to equilibrium approach }}

Finally, we are in a position to address the main claim of this work;  diffusive particle systems, within the framework of the MFT belong either the Arrhenius universality class, or the excluded volume universality class. In this section, we present the proof. 

Importantly, the proof provided here has to be carried out independently for each dynamics $D,\chi$. Therefore, we start by stating the essence of the proof, and only then continue to demonstrate it for: non-interacting particles, the simple exclusion process, the strong particles model, the simple inclusion process and finally the family of zero-range processes.

The premise of our method is to notice that at the $\Delta U \gg D_0$ limit, $\Phi \ll 1$. Namely, the escape rate of particles from a deep trap is small. Considering an equilibrium density profile $\rho_{\rm{eq}}$ (correspondingly $F_{\rm{eq}}$), where $J(\rho_{\rm{eq}})=0$ (correspondingly $J_{F_{\rm{eq}}}=0$)  implies $\Phi=0$ directly from \eqref{eq:mini Phi F}. 
The equilibrium density profile is small but non vanishing at the absorbing boundary, in contradiction with the boundary conditions \eqref{eq:rho BCs}. Therefore, it seems plausible that the optimal fluctuation can be written as a perturbative expansion around the equilibrium fluctuation. Let us construct this perturbation theory.



\subsection{The essence of the proof
\label{subsec: proof essence}}

 In order to estimate the escape rate $\Phi$, we need to solve the Euler-Lagrange equation \eqref{eq:EL equation F} subjected to the boundary conditions \eqref{eq:rho BCs}. The premise of the method is that the optimal profile $F$ can be approximated by invoking a perturbative analysis around the equilibrium setup
\begin{align}
    F= F_{\rm{eq}}+\omega \delta F + O(\omega^2),
\label{eq:perturbation}
\end{align}
where $\omega$ is a small parameter, yet to be determined, and $\delta F$ is an order $1$ perturbation. It is important to state from the get-go that it is hard to carry out the perturbation theory to the next order. Nevertheless, it will prove useful for the task at hand, as $\Phi \asymp \omega^2 $ to leading order.


First, we explicitly find $F_{\rm {eq}}$ as the solution of $J_{F_{\rm{eq}}}=0$. The equilibrium solution $F_{\rm {eq}}$ naturally depends on the particle mass in the system $M = L \overline{\rho}_0$. Therefore $F_{\rm{eq}}$ is determined from $J_{F_{\rm{eq}}}=0$ up to a constant we denote by $A$. Then, it is important to find $A = A(\frac{\Delta U}{D_0},\overline{\rho}_0)$. We expect this is typically possible analytically at the limit of large $\Delta U/D_0$. 

 $F_{\rm {eq}}$,  the equilibrium component, satisfies the reflecting boundary condition, but it does not account for the absorbing boundary condition. We posit that the correction term $\omega \delta F$ is responsible to mend the absorbing boundary condition. Namely, $F_{\rm{eq}}+\omega \delta F |_{x=1}=0$. More precisely, we fix $\delta F|_{x=1}=-1$ so that
 \begin{equation}
 \label{eq:fixing omega}
 \omega = F_{\rm{eq}}|_{x=1} .    
 \end{equation}
  This implies that to leading order in $\omega$
\begin{equation}
\label{eq:perturbative minimization}
    \Phi = L D_0\omega^2 \min_{\delta F} \int dx \, (\partial_x \delta F + \frac{\chi'}{4D_0 \chi^{1/2}}\partial_x U \delta F )^2, 
\end{equation}
where $\chi= \chi(F_{\rm{eq}}) $.  
 The new minimization problem of $\delta F$ is in no way simpler than the original problem and we make no attempt to solve it. Instead, we notice $\delta F$ is of order $1$ and therefore the integral in \eqref{eq:perturbative minimization} is typically non-vanishing and leads to a polynomial function of $\Delta U / D_0$.  Therefore, it is $\omega$ that determines the exponential decay of $\Phi$. 
 
 In simple words, using this perturbative approach, we have thus reduced the problem from solving a hard minimization problem on $F$, to finding the proper scaling $\omega$. Finally, after determining $\omega$, we recover $\Phi \asymp \omega^2$ to recover the generalized AL at the limit where $\Delta U/D_0 \gg 1$. 

In what follows, we implement this methodology for a set of paradigmatic lattice gas models. 


\subsection{Non-interacting particles}

We first focus on non-interacting particles. It has long since been established that the AL is satisfied in this case. In terms of \eqref{gen-AL}, it corresponds to $g=1$, independent on the density $\overline{\rho}_0$. Keeping the AL result in mind, we revisit the escape problem for non-interacting particles using our new methodology in order to re-affirm $g=1$.

Within the MFT formalism, $D=1$, and $\chi=\rho$ for non-interacting diffusive particles. In this case, the $F$ transformation \eqref{eq:F transformation} implies  $\rho=F^2$. With the above in mind, we write the escape rate for non-interacting particles $\Phi_{NI}$, the resulting Euler-Lagrange equation, and the boundary conditions 
\begin{eqnarray}
    \Phi_{\rm{NI}} &=& D_{0}L \int dx (\partial_x F + \frac{1}{2D_0} F \partial_x U)^2 + \Lambda(F^2 - \overline{\rho}_0 ),
\nonumber \\  
&& \hat{H}_{FP} F = \Lambda F,
\\ \nonumber 
&& \partial_x F + \frac{1}{2D_0} F \partial_x U |_{x=0} = F|_{x=1} = 0. 
\end{eqnarray}
According to the recipe of Sec.~\ref{subsec: proof essence}, we posit \eqref{eq:perturbation}, 
and for a vanishing $J_{F_{\rm{eq}}}$, we find
\begin{eqnarray}
\label{eq: NI equilibrium sol}
    F_{\rm{eq}} &=& A \, \rm{e}^{-\frac{U(x)}{2D_0} }, \quad \rm{and}
    \nonumber  \\ 
    A^2 &=& \frac{\overline{\rho}_0 }{ \int dx \, \rm{e}^{-U(x)/D_0}},
\end{eqnarray}
where $A$ is obtained from the mass conservation at the leading order. At the large $\Delta U/D_0$ limit, we can explicitly evaluate $A^2 = \overline{\rho}_0 \frac{\partial_x U(x=0)}{D_0} + O(1)$, 
provided the first derivative around $x=0$ is non-vanishing. More generally, when $\partial^k _x U(x=0) =0 $ for $k<n$, one finds $A^2 \propto \overline{\rho}_0 (\Delta U /D_0 )^{1/n} $ up to subleading contributions in $\Delta U/D_0$.


From \eqref{eq:fixing omega}, satisfying the absorbing boundary condition, we find
\begin{equation}\label{NI_omega}
    \omega = A \, \rm{e}^{-\frac{\Delta U}{2D_0}} . 
\end{equation}
This choice is justified if $\omega$ is indeed small. Since $A$ is polynomial in $\Delta U /D_0$, and $\overline{\rho}_0$ is finite, we confirm $\omega \ll 1$ at large $\Delta U /D_0$ and the perturbation theory is justified. Moreover, we verify that 
\begin{eqnarray}
    \Phi_{NI} &=& D_0 L \omega^2 \rm{Poly}(\Delta U /D_0) 
    \\ \nonumber
    &=& D_0  M \rm{Poly}(\frac{\Delta U}{D_0} ) \rm{e}^{-\frac{\Delta U}{D_0}},
\end{eqnarray}
where the polynomial contribution comes from both the $\omega^2$ term and from the integral in \eqref{eq:perturbative minimization}. Therefore, we have reconstructed the famous AL for non-interacting particles. In what follows, we show that the same methodology, with a tad more technicality, re-affirms the classification to two AL universality classes; differentiating between processes with and without excluded volume. 


\subsection{The simple exclusion process (SEP)}

In this subsection, we focus on inferring $g$ of \eqref{gen-AL} for the the simple exclusion process (SEP). It will demonstrate that the SEP belongs to the excluded volume universality class, as predicted by \eqref{gen-AL}.

The SEP is a lattice gas model, where each site is either occupied or not. A particle at site $i$, jumps to the neighboring site $j$ with rate $n_i (1- n_j) $ where $n_{i}$ states the onsite occupancies (see Fig.~\ref{fig:main}). The SEP is a paradigm in the study of nonequilibrium as it supports a Bethe Ansatz solution \cite{Derrida07,derrida2004current}. Moreover, it displays genuine nonequilibrium behaviour like long-range correlations at nonequilibrium steady state, a steady state measure which is not a product measure, dynamical phase transitions and more \cite{mallick2015exclusion,shpielberg2018universality}. Physically, the SEP was used to various biophysical processes, e.g. dynamics of Ribosomes along RNA, molecular motors. Further applications include transport through a series of quantum dots, and traffic flows \cite{chou2011non}. For our purpose, the SEP is the simplest model to study the breakdown of the AL universality class and thus revealing the excluded volume universality class.

Within the MFT formalism, the SEP is defined by $D=1$, and , $\chi=\rho(1-\rho)$. The $F$ transformation \eqref{eq:F transformation} implies $\rho =\sin^2  F $ for the SEP. Let us, once again,  collect the escape rate $\Phi_{\rm{SEP}}$ and the boundary conditions for the SEP \footnote{Here we refrain from writing the Euler-Lagrange equation as the scaling of $\omega$ stems from the absorbing boundary condition.  }
\begin{eqnarray}
    \frac{\Phi_{\rm{SEP}}}{D_0 L} &=&  \min_F \int dx \, J^2 _F + \Lambda (\sin^2 F - \overline{\rho}_0),  
    \nonumber \\  
      F|_{x=1}  &=& \partial_x F  + \frac{1}{4D_0}\sin 2F \partial_x U|_{x=0} = 0 , 
      \nonumber \\ 
       J_F &=& \partial_x F + \frac{1}{4D_0} \sin 2F \partial_x U  .
\end{eqnarray}
 Our methodology states that we need to find $F_{\rm{eq}}$  explicitly.   Imposing $J_{F_{\rm{eq}}}=0$ $\forall x$ implies 
 \begin{equation}\label{sep_feq}
     F_{\rm{eq}} = \arctan (A \, \rm{e}^{-U/2D_0}). 
  \end{equation}
$A$ is inferred from the mass conservation $\overline{\rho}_0 = \int dx \, \sin^2 F_{\rm{eq}} $, i.e. 
\begin{equation}
\label{eq:SSEP density A}
    1- \overline{\rho}_0 = \int dx \, \frac{1}{1+ A^2 \rm{e}^{-U(x)/D_0}}. 
\end{equation}
At large $\Delta U/D_0$ values, and assuming monotonous potentials, it is possible to evaluate $A^2 = \rm{e}^{U(\overline{\rho}_0)/D_0}$, up to subleading corrections. Indeed, we notice that \footnote{The subleading term $D_0/\Delta U$ can change to a different power law if $\partial_x U|_{x=0} $ vanishes.}:
\begin{equation} \label{eq: Messy integral SSEP}
\begin{split}
\int   \frac{dx}{1+ A^2 \rm{e}^{-U(x)/D_0}}   & = \left(\int^{\overline{\rho}_0} _0 + \int^1  _{\overline{\rho}_0}  \right) 
     \frac{dx}{1+  \rm{e}^{{
    \frac{U(\overline{\rho}_0)-U(x)}{D_0}}} }  
\\
 &  \approx O( \frac{D_0}{\Delta U} ) + 1- \overline{\rho}_0. 
\end{split}
\end{equation}
Now that we have evaluated $A$, we recall that \eqref{eq:fixing omega} implies 
\begin{eqnarray}\label{sep_omega}
\omega &=& F_{\rm{eq}}|_{x=1} = \arctan{A \rm{e}^{-\Delta U /2D_0}}    ,
\\ \nonumber 
\Phi_{\rm{SEP}}    &=& D_0 L \rm{Poly}(\Delta U/ D_0)\omega^2  
\\ \nonumber 
&& = D_0 L \rm{Poly}(\Delta U/ D_0) \rm{e}^{-\Delta U  g / D_0 } , 
\end{eqnarray}
for deep potentials, and with $g=1-U(\overline{\rho}_0)/\Delta U$ for monotonous $U$ as suggested in \cite{kumar2023arrhenius}. Note that $\delta F$ is of order 1 and thus the polynomial contributions arise from the integral. 
 Moreover, for a finite density $\omega^2 = \arctan  \rm{e}^{-\Delta U g / D_0  } \ll 1$, thus justifying the perturbative approach.

We have  shown that this perturbative scheme is sufficient to capture the generalized Arrhenius law in monotonous potentials. The extension to non-monotonous potentials can be pursued by the perturbative scheme and amounts to re-evaluating the integral  \eqref{eq: Messy integral SSEP}. While non-monotonous potentials lead to interesting results on their own right, a focused study will be presented in \cite{Kumar2024Nonmonotone}.

We have verified for the SEP the main claim of \eqref{gen-AL}. The excluded volume results in $g<1$. We note that the limit of vanishing density where $g=1$ corresponds to the reflected Brownian motion process \cite{jepsen1965dynamics,barkai2010diffusion,krapivsky2010kinetic}. 
Unlike the finite density SEP, the reflective Brownian motion introduces point particles that cannot overtake one another. The escape problem is indifferent to tagged particles. Therefore, as expected, $g=1$ in the vanishing density limit. In summary, the effect of excluded volume is to decrease $g<1$ as predicted by \eqref{gen-AL} for finite densities $\overline{\rho}_0$.


In what follows, we show that even in the case of a more complex dynamics than the SEP, involving excluded volume effects, we still recover the same $g$ given by \eqref{gen-AL}. Namely, processes in the excluded volume universality class lead to the same $g$, independent of the dynamics.


\subsection{The strong particles model}

In this section we consider a class of lattice gas models known as strong particles \cite{gabrielli2018gradient}. Once more, each lattice site is either occupied or not. Moreover, chunks of $k$ occupied sites can move in their entirety with rate $r(k)$ one step to the right or left. One can envision that the particles are strong enough to push a whole line of particles, hence the name (see Fig.~\ref{fig:main}). In the literature this processes was sometimes referred to as the long range jumping model for obvious reasons \cite{shpielberg2017numerical}.

The strong particles model is a gradient type model. This implies an analytic calculation of the diffusivity \cite{spohn2012large,gabrielli2018gradient}. In our case, 
\begin{equation}
    D = \sum^{\infty} _{k=1} \,  k^2 r(k) \rho^{k-1}.
\end{equation}
As long as the rates allow all possible configurations, we have the free energy density $f(\rho)= \rho \log \rho + (1-\rho)\log (1-\rho)$. This implies that the mobility is given by 
\begin{equation}
    \chi = \rho(1-\rho) D.
\end{equation}

Clearly, for $r(k=1)=1$ and $r(k>1)=0$ we reduce to the SEP. Here, we consider the rate $r(k)=1/k^2$, leading to $D=1/(1-\rho) $ with $\chi = \rho$. This simple form of $D,\chi$ allows to perform the $F$ transformation \eqref{eq:F transformation}. We find $\rho = \chi = \tanh^2 F$ with $D=\cosh^2 F $.

Collecting the setup for the strong particles model, we have 
\begin{eqnarray}
    \frac{\Phi_{\rm{SPM}}}{D_0 L} &=&  \min_F \int dx \, J^2 _F + \Lambda (\tanh^2 F - \overline{\rho}_0),  
    \nonumber \\  
      F|_{x=1}  &=& \partial_x F  + \frac{1}{2D_0}\tanh F \partial_x U|_{x=0} = 0 , 
      \nonumber \\ 
       J_F &=& \partial_x F + \frac{1}{2D_0} \tanh F\partial_x U  .
\end{eqnarray}
Setting $J_{F_{eq}}=0$, we find $F_{eq} = \arcsinh A \rm{e}^{-U/2D_0}$, where $A$ is determined by mass conservation
\begin{equation}
\label{eq:mass conservation strong particles}
    \overline{\rho}_0 = \int dx \tanh^2 F_{eq} = \int dx \frac{1}{1+A^{-2} \rm{e}^{U(x)/D_0} }. 
\end{equation}
For large $\Delta U/D_0$, and for monotonous potentials, we can evaluate $A^2 = \rm{e}^{U(\overline{\rho}_0) / D_0}$.


Now, we have from \eqref{eq:fixing omega} that $\omega =  \arcsinh A \rm{e}^{-\Delta U/2D_0}$, leading to the escape rate for deep potentials 
\begin{equation}
    \Phi_{\rm{SPM}} = D_0 L \rm{Poly}
(\Delta U /D_0)  \rm{e}^{-\Delta U g /D_0},
\end{equation}
where $g = 1 - U(\overline{\rho}_0) / \Delta U$ as expected. To conclude, $g$ is identical for the SEP and the strong particles model. They both belong to the excluded volume universality class.



\begin{figure*}
\centering
\includegraphics[height=7cm]{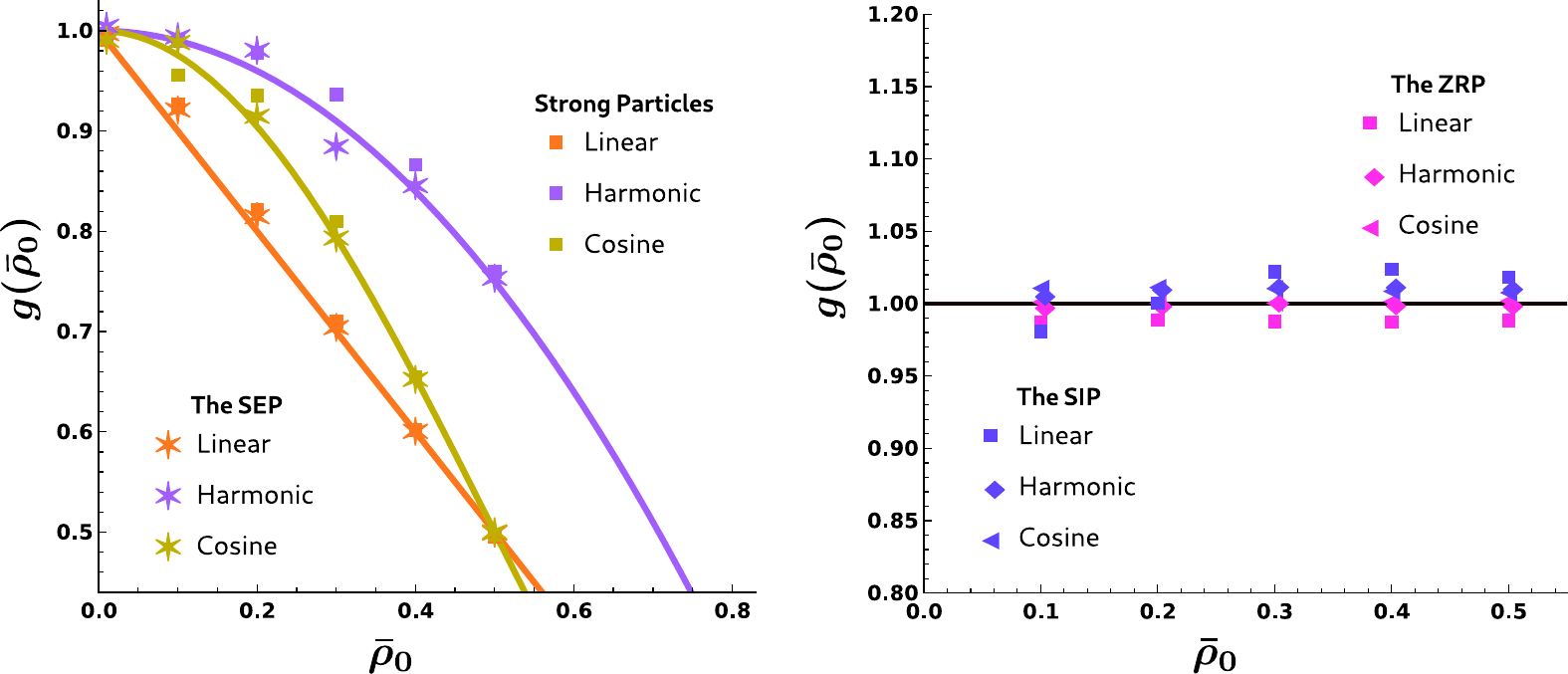}
\caption{
Numerically solving \eqref{eq:mini Phi F} subjected to  \eqref{eq:EL equation F}, the two figures present $g$ as a function of the particle density in the trap for three potentials: The linear potential $U/\Delta U = x$, the Harmonic potential $U/\Delta U = x^2$ and the Cosine potential $U/\Delta U = \frac{1}{2}(1- \cos \pi x) $. The left plot presents $g$ for the SEP and the strong particles model, both falling within the excluded volume universality class. For the monotonous potentials under study, $g = 1- U(\overline{\rho}_0)/\Delta U$ as obtained from the minimum energy configuration. In the right plot, $g$ is presented for the inclusion process and for a ZRP with $\chi = \rho(1+\rho)$. It is confirmed that both processes belong to the Arrhenius universality class, with $g=1$. For both plots, the numerical values show good agreement with the theoretical prediction and are within the expected errors of the numerical method.    
} 
\label{fig:strong particle}
\end{figure*}

 We verify numerically that the SEP and strong particles model both belong to the excluded volume universality class. See  Fig.~\ref{fig:strong particle}. 

We have established the excluded volume universality class, where $g$ is independent of the type of interactions. 
We move on-wards to analyze the simple inclusion process (SIP), which involves attractive interactions, but no excluded volume effects. This, in order to argue that even with interactions, but no excluded volume effects, the process lies within the Arrhenius universality.

\subsection{The simple inclusion process (SIP)}

Here, we analyze the simple inclusion process (SIP). The SIP is a paradigm model for attractive interactions. Naively, one might have expected that since the repulsive interactions of the SEP led to $g<1$, the attractive interactions of the SIP lead to $g>1$. We will prove here this is not the case and that for the SIP $g=1$ for any particle density $\overline{\rho}_0$, as is the result of \eqref{gen-AL}. In short, the SIP belongs to the Arrhenius universality class. 

The SIP is a lattice gas model, where the particle occupancy at each site is unbounded. A particle at site $x$ jumps to a nearest neighboring site $y$ with rate $n_x (1+n_y)$ where $n_{x,y} $ are the particle occupancies \cite{reuveni2012asymmetric,appert2008universal,shpielberg2021thermodynamic,baek2016extreme} (see Fig.~\ref{fig:main}). Namely, where the SEP is the classical analog of Fermions, the SIP is the classical analog of Bosons. It is often the case where if the SEP dynamics can be solved analytically, so can the corresponding SIP dynamics \cite{appert2008universal}.

 In the literature, one can find numerous studies on the SIP as a model that facilitates condensation \cite{grosskinsky2011condensation,grosskinsky2013dynamics}. Here, we focus on diffusive transport, and thus avoid the condensation region. Hence, the SIP densities are assumed low enough to avoid any condensation transition. For $\overline{\rho}_0<0.6$, the absence of condensation is numerically verified.

Within the MFT framework, the SIP is captured by $D=1$, and $\chi=\rho(1+\rho)$. The $F$ transformation \eqref{eq:F transformation} reads $\rho = \sinh^2 F$. We therefore recover 

\begin{eqnarray}
   \frac{ \Phi_{\rm{SIP}} } {D_0 L} &=&  \min_F \int dx \, J^2 _F  + \Lambda (\sinh^2 F - \overline{\rho}_0), 
   \\ \nonumber
   F|_{x=1} &=& \partial_x F+ \frac{1}{4D_0}\sinh 2F \partial_x U |_{x=0} =0,  
   \\ \nonumber
   J_F &=&  \partial_x F+\frac{1}{4D_0} \sinh 2F \partial_x U.
\end{eqnarray}

Setting $J_{F_{\rm{eq}}}=0$, we find 
$F_{\rm{eq}}= \text{arctanh} A \rm{e}^{- U/2D_0 } $. This implies, together with \eqref{eq:fixing omega}
that
\begin{eqnarray}\label{sip_omega}
\omega &=& \text{arctanh} \, A \rm{e}^{-\Delta U/2D_0} ,     \\ \nonumber 
1+ \overline{\rho}_0 &=& \int dx \, \frac{1}{1-A^2 \rm{e}^{-U(x)/D_0}}, 
\end{eqnarray}
where we have recovered $A$ from the mass conservation.  
Here, we can immediately notice that $A^2<1$ keeping the integral convergent, and that for finite $\overline{\rho}_0$ $0<A^2<1$ is finite when $\Delta U/D_0 \gg 1$.  
Evaluating the latter integral can be done exactly only for a set of potentials. Nevertheless,  for $\overline{\rho}_0 \rightarrow 0$ we have $A^2 \rightarrow 0$ and similarly as $\overline{\rho}_0 \rightarrow \infty $ we have $A^2 \rightarrow 1^-$. Moreover, $A^2$ is an increasing (and bounded) function of $\overline{\rho}_0$. Therefore, we find $\omega^2 = \text{arctanh}^2 (A \rm{e}^{-\Delta U/2D_0 }) \approx A^2 \rm{e}^{-\Delta U /D_0 }$. This implies 
\begin{equation}
    \Phi_{\rm{SIP}} =D_0 L \rm{Poly}(\Delta U /D_0) \rm{e}^{-\Delta U/D_0},
\end{equation}
following \eqref{gen-AL}, and as proposed in \cite{kumar2023arrhenius}. The SIP, lacking excluded volume effects, is indeed within the Arrhenius universality class.

Let us conclude the analysis with the zero-range processes, allowing to account for non-trivial interactions, repulsive and attractive, but with no excluded volume effects.  

\subsection{The zero-range processes (ZRP)}

In this subsection, we turn to study the set of zero-range processes (ZRP). Unlike the previous models, where the interaction was restricted to either repulsive or attractive, the set of ZRP allow to have repulsive, attractive or a mixture of repulsive and attractive interactions. Like the SIP, the site occupation is unbounded. Therefore, there are no excluded volume effects, and \eqref{gen-AL} implies $g=1$. Here, we show this is indeed the case, showing that the type of interaction plays no role in the determination of $g$, where no excluded volume effects are present in lattice gas models.

The ZRPs are a set of processes that describe lattice models where the jump rate for a particle, to jump from site $k$ to a neighboring site is given by $f(n_k)$, where $n_k$ is the local occupancy at site $k$. Here, $f(n)$ is an arbitrary non-negative function, where we fix $f(0)=0$ and assume $f(n)>0$ for $n>0$ (see Fig.~\ref{fig:main}
). 
Macroscopically, the ZRP is captured by $D=\partial_\rho \chi $ with $\chi$ an increasing function of the density and with the natural condition $\chi(0)=0$ \cite{levine2005zero}. $\chi$ is a non-trivial function of $f$, except for $f(n)=n$, corresponding to $\chi=\rho,D=1$, where we recover the non-interacting particles model.

The $F$ transformation implies $\chi =\rho = F^2$. Hence, we have 
\begin{eqnarray}
    \frac{\Phi_{\rm{ZRP}}}{D_0 L}
 &= &  \min_F \int dx \, 
 J^2 _F 
+\Lambda( \rm{inv}[\chi] -\overline{\rho}_0) ,
 \nonumber  \\ 
  F|_{x=1} &=&  \partial_x F + \frac{1}{2D_0}\chi^{1/2} \partial_x U|_{x=0}  =0, 
 \\ \nonumber 
  J_F  &=& 
 \partial_x F + \frac{1}{2D_0}\chi^{1/2} \partial_x U. 
\end{eqnarray}
Here we have denoted $\rm{inv}[\chi]$ as the inverse function to $\chi$.  Requiring $J_{F_{\rm{eq}}}=0$  implies  $F_{eq} = A \rm{e}^{-U/2D_0}$, where the constant $A$ is found from the mass conservation 
\begin{equation}
    \overline{\rho}_0 = \int dx \, \rm{inv}[\chi]  (F^2 _{eq}).
\end{equation}
From \eqref{eq:fixing omega}, we find $\omega = A \rm{e}^{-\Delta U/D_0}$. We stress that the perturbative approach works as long as $\omega\ll 1$. Hence, we will need to infer the value of $A$, which of course depends on the mobility $\chi$.

Since $\chi$ is a monotonous increasing function of $\rho$, $\rm{inv}[\chi]$ is also a monotonous increasing function of $F$. Hence, together with $\rm{inv}[\chi] (0)=0$, we can use the mass conservation equation to express $A$
\begin{equation}
    \overline{\rho}_0 = \int dx \, \rm{inv}[\chi](A^2 \rm{e}^{-U/D_0} ) \approx c \frac{D_0}{\Delta U} \rm{inv}\left[\chi \right] \left(A^2 \right), 
\end{equation}
where $c>0$ is a constant depending on the exact shape of $\chi$. Taking this into account we demand \begin{equation}
\label{eq:ZRP omega small}
    \omega^2 = \chi \left(\frac{1}{c D_0 }\Delta U \overline{\rho}_0 \right) \rm{e}^{-\Delta U/D_0} \ll 1, 
\end{equation}
for the validity of the perturbation approach. Eq.~\eqref{eq:ZRP omega small} is satisfied for a polynomial $\chi$ and we find as expected
\begin{equation}
    \Phi_{\rm{ZRP}} \asymp \rm{e}^{-\Delta U/D_0}.
\end{equation}
In the rather extreme case, where $\chi$ is an exponentially increasing function, there exists some density where $\omega$ is not longer small. In this case, the perturbative approach breaks down. Physically, an exponentially increasing mobility and diffusivity imply that particles do not tend to bunch up, and that the particle jump rate is increasing with the local occupancy. In this case, particles leave the trap fast. It is beyond the scope of the present work to asses what is the escape rate in such a scenario. Nevertheless, notice that this is the exact opposite case to one leading to a condensation transition \cite{kafri2002criterion}, where the particle rates become slow when the onsite density is large.

\begin{widetext}
    \renewcommand{\arraystretch}{3}
\begin{table*}
\centering
\caption{Summary of the Generalized AL
}

\begin{tabular}{||m{2.6cm}|m{1.7cm}|m{2.6cm}|c|c||c}

\cline{1-5}
\textbf{Models} & \hspace{0.4cm}\bm{$D$ \&\ $\chi$} &\hspace{1cm}\bm{ $F_{eq}$} &  \bm{$\omega$} & \textbf{AL, }\bm{$\Phi\asymp e^{-\frac{\Delta U}{D_0}g(\bar{\rho}_0)}$ }\\
\cline{1-5}
\textbf{Non-interacting particles} & \parbox{1.7cm}{$D=1$\\ $\chi=\rho$}  & \hspace{0.45cm}$A e^{-\frac{ U(x)}{2 D_0}}$ (\ref{eq: NI equilibrium sol}) & $A e^{-\frac{\Delta U}{2 D_0}}$ (\ref{NI_omega}) & $g=1$ & \rdelim\}{3}{20.5mm}[\parbox{12 mm}{\textbf{\parbox{1.5cm}{Arrhenius\\universality}}}]\\
\cline{1-5}

\textbf{SIP} & \parbox{1.7cm}{$D=1$\\ $\chi=\rho(1+\rho)$} &  arctanh$(A e^{-\frac{ U(x)}{2 D_0}})$ & arctanh$(A e^{-\frac{\Delta U}{2 D_0}})$ (\ref{sip_omega}) &$g=1$ \\ 
\cline{1-5}

\textbf{ZRP} & \hspace{0.2cm}$D=\partial_{\rho}\chi$ &  \hspace{0.45cm}$A e^{-\frac{ U(x)}{2 D_0}}$  &$A e^{-\frac{\Delta U}{2 D_0}}$ & $g=1$ \\
\cline{1-5}

\textbf{SEP} & \parbox{1.7cm}{$D=1$ \\ $\chi=\rho(1-\rho)$} &  arctan$(A e^{-\frac{ U(x)}{2 D_0}})$ (\ref{sep_feq})& arctan$(A e^{-\frac{\Delta U}{2 D_0}})$ (\ref{sep_omega}) & $g=1-\frac{U(\bar{\rho}_0)}{\Delta U}$ 
 & \rdelim\}{2}{20.5mm}[\parbox{12 mm}{\textbf{\parbox{3cm}{Excluded volume\\universality}}}]\\
\cline{1-5}

\textbf{Strong particles model} & \parbox{1.7cm}{$D=\frac{1}{1-\rho}$\\ $\chi=\rho$} &  arcsinh$(A e^{-\frac{ U(x)}{2 D_0}})$ &arcsinh$(A e^{-\frac{\Delta U}{2 D_0}})$ & $g=1-\frac{U(\bar{\rho}_0)}{\Delta U}$\\
\cline{1-5}

\end{tabular}
\end{table*} 
\end{widetext}

\section{Discussion
\label{sec:discussion}}


In this work, we have delved deeper into the Arrhenius Law for interacting particle systems. To this end, we have computed the survival probability of  interacting diffusive particles in a deep potential trap. 
Typically, the survival time is significantly longer than the diffusive time $L^2 / D_0$ when the trap is deep, i.e. $\Delta U /D_0 \gg 1$. In this case, the survival probability takes the large deviation form  $S(t) \asymp \rm{e}^{-t \Phi}$, where $\Phi$ is the large deviation function. To asses $\Phi$ -- the escape rate from the trap -- we have employed the MFT, a nonequilibrium hydrodynamic theory for diffusive systems. Analytically, we have established, for a list of paradigmatic processes, that the escape rate falls into one of the two universality classes. If particle occupancy is unbounded, the process belongs to the Arrhenius universality class, where $\Phi \asymp \rm{e}^{-\Delta U / D_0}$. This is demonstrated for the non-interacting case, the SIP and a large class of zero-range processes. If particle occupancy is bounded, i.e. particles generate excluded volume, the process belongs to the excluded volume universality and $\Phi \asymp \rm{e}^{-\Delta U g /D_0}$, where $g = 1- U(\overline{\rho}_0)/\Delta U$ for monotonous potentials. Importantly, $g$ is independent of the dynamics of process. This was demonstrated for the SEP and the strong particles model which share the same $g$ despite having  completely different dynamics, see Fig.~\ref{fig:strong particle}. We refer to Table I for a concise summary of the generalized AL.

An intuitive way to interpret $g$ in the escape rate is through the minimum energy configuration \cite{kumar2023arrhenius}. At low temperatures, i.e. $D_0\ll \Delta U$, an equilibrium system is dominated by its minimum energy configuration. Hence, the activation energy $\Delta U g$ is the necessary energy to activate a single particle. For deep potentials, the minimum energy configuration packs the particles as tightly as possible around the global minimum. Particles that can be packed into a single point at finite energy will be found in the Arrhenius law. Particles with exclusion dynamics change their minimum energy configuration as a function of their density in the trap. For a monotonous trap, we expect to find the highest energy particle at $x=\overline{\rho}_0$. Hence, the activation energy is $\Delta U g = \Delta U - U(\overline{\rho}_0)$. Indeed this intuitive argument, suggested in \cite{langer1969statistical} for systems with linear dynamics, allows to infer the rigorous results obtained in this work.

The approach developed here significantly reduces the complexity of calculating the escape rate $\Phi$. Previously in \cite{kumar2023arrhenius}, $\Phi$ was obtained by direct solution of the non-linear Euler-Lagrange equation \eqref{eq:EL equation F}. Analytically, solving  \eqref{eq:EL equation F} is feasible for a limited set of potentials. Numerically, solving \eqref{eq:EL equation F} via a shooting method is restricted to $\Delta U / D_0 $ no larger than $\sim 40$. Using the perturbative approach developed here, we notice that at the expanse of ignoring the sub-exponential terms of the escape rate, the perturbative approach allows to gain analytical insight for any potential. Numerically, one can employ the perturbative approach which basically boils down to extracting $A$ from the mass term, see e.g. \eqref{eq:SSEP density A}. For the SEP, one can easily go to $\Delta U / D_0   \sim 100$ by numerical integration of  \eqref{eq:SSEP density A}. The numerical advantage of the perturbative approach will be crucial when studying non-monotonous potentials. This will be discussed at length in a subsequent publication \cite{Kumar2024Nonmonotone}.

The perturbative method is able to corroborate for a number of important models
the existence of the two universality classes; the Arrhenius universality class for lattice gas models with unbounded particle occupancy, and the excluded volume universality class for lattice gas models with bounded particle occupancy. One case where the perturbative approach is inconclusive, is for a ZRP where   $\chi$ is exponentially increasing for large densities. This process, microscopically, corresponds to a model where particle bunching leads to strong repulsion. While the perturbative approach fails to capture the escape rate $\Phi$, it should still be possible to employ the shooting-method algorithm in \cite{kumar2023arrhenius} to understand whether the universality is maintained. Moreover, it could also be possible to invoke microscopic techniques in order to capture the escape rate $\Phi$ \cite{evans2005nonequilibrium}.

Another point worth discussing is the importance of the  $F$ transformation in the analytical analysis. First, there exist non-particle transport models, like the KMP \cite{kipnis1982heat}, where a continuous quantity like energy is being transported. 
For the KMP model, with $D=1,\chi=\rho^2$, the $F$ transformation diverges. Simply put, this implies,  that the large deviation form of the survival probability can no longer be assumed. We do not expect such escape problem could be successfully analyzed using the MFT. Second, the macroscopic presentation using $D,\chi$ of many particle models is either unknown, or given by convoluted expressions \cite{arita2014generalized}. This implies that only numerical approximations to the $F$ transformation are possible. It would be useful to develop such a protocol to advance the perturbative approach analysis to deal with arbitrary diffusive dynamics.

To conclude, 
the successful identification of the two universality classes motivates further studies, beyond the diffusive regime. Can a similar classification in active systems, multi-particle species, particles with inertia be observed? Is it possible to identify other universality classes when we leave the realm of lattice gas models, e.g. in soft particles? These questions show that indeed the study of the escape problem from a deep trap is far from resolved.

\begin{acknowledgements}
OS and AP thank Eli Barkai for pointing out the reflective Brownian motion problem. OS acknowledges the support of the Erwin Schr\"{o}dinger International Institute for Mathematics and Physics during the thematic program DPS22. AP gratefully acknowledges research support from the Department of Science and Technology, India, SERB Start-up Research Grant Number SRG/2022/000080 and Department of Atomic Energy, Government of India.
\end{acknowledgements}


\twocolumngrid

\bibliography{main}

\end{document}